\documentstyle[preprint,aps,epsfig]{revtex}
\tightenlines

\def\be{\begin{equation}}
\def\ee{\end{equation}}
\def\ep{\epsilon}
\def\bee{\begin{eqnarray}}
\def\eee{\end{eqnarray}}
\def\arf{\alpha}

\begin{document}
\draft

\title{Ground state of a polydisperse electrorheological solid: \\
 Beyond the dipole approximation}
\author{H. Sun$^{a,b}$, K. W. Yu$^a$}
\address{$^a$Department of Physics, the Chinese University of Hong Kong, 
Shatin, NT, Hong Kong \\
 $^b$Department of Physics, Suzhou University, Suzhou 215006, China}

\maketitle

\begin{abstract}
The ground state of an electrorheological (ER) fluid has been studied 
based on our recently proposed dipole-induced dipole (DID) model.
We obtained an analytic expression of the interaction between chains of 
particles which are of the same or different dielectric constants. 
The effects of dielectric constants on the structure formation in 
monodisperse and polydisperse electrorheological fluids are studied in a 
wide range of dielectric contrasts between the particles and the base 
fluid. Our results showed that the established body-centered tetragonal 
ground state in monodisperse ER fluids may become unstable due to a 
polydispersity in the particle dielectric constants. 
While our results agree with that of the fully multipole theory, the DID 
model is much simpler, which offers a basis for computer simulations in 
polydisperse ER fluids.
\end{abstract}

\section{Introduction}

The study of structure formation of electrorheological (ER) fluids has 
attracted increasing interest in recent years for its fundamental and 
technological values. Upon the application of an external electric field, 
the suspended particles in an ER fluid aggregate into chains 
and then columns parallel to the field, and drastically change the 
rheology of the suspension \cite{win}. Tao and his coworker first 
suggested 
the existence of microcrystalline structures inside the columns and 
identified its ground state to be a body-centered tetragonal (bct) 
lattice \cite{Tao1}. These authors developed an analytic method based on 
a 
point-dipole (PD) approximation, 
i.e., treating the dielectric spheres as point dipoles interacting with
one another.
The possible ground state is the configuration that minimizes the 
dipole interaction energy and consequently the total Coulomb energy. 
The idea was soon confirmed by computer simulations and experiments 
\cite{tao2,chen}. The discovery of this property is not only helpful to 
understand further the ER response to the external field, but also offers 
a new technique to form mesocrystals with unique photonic properties 
\cite{wen99,golo}.

After many efforts to reveal the details of structure formation in an ER 
fluid, it has been known that what structure will be formed in such a 
system actually depends on many factors, such as volume fractions 
\cite{Mart}, the size distribution of particles \cite{ota,wu}, the field 
frequency \cite{wen00}, etc. Combining the electrorheological and 
magnetorheological effects \cite{wen99,tao98} or applying a rotating 
electric field \cite{mar2} was also found to cause a structure transition 
from the bct to face-centered cubic (fcc) lattices \cite{lo}.  

The influence of the dielectric constant is another interesting topic 
since dielectric mismatch between the particles and the fluid is widely 
accepted as the main reason for the ER phenomenon. Davis found that the 
bct, fcc and hexagonal closed packed (hcp) lattices degenerate when the 
particle permittivity ($\ep_p$) is much larger than that of the base 
fluid ($\ep_f$) \cite{dav}. Clercx and Bossis \cite{cle} developed a 
fully multipolar 
treatment and compared the bct and fcc lattices for some values of 
$\ep_p/\ep_f$. Their results are in agreement with Davis. 
Lukkarinen 
and Kaski studied both the free energies of many types of lattices and 
dynamical 
effects in an ER fluid containing particles with the dielectric constant
being greater and/or smaller than that of the fluid \cite{luk98,luk01}. 
However, all of the previous work has either limited to the extreme case 
($\ep_p\gg\ep_f$) or been too complicated to be adopted in further 
studies. 
And the understanding of the way dielectric constant affects structure 
formation is far from being sufficient.

Apart from theoretical work, computer simulation is another effective 
method to gain insight into ER effects. The point-dipole approximation is 
routinely adopted in simulation \cite{kli} because of its simplicity. 
Since many-body and multipolar interactions are neglected in this 
approximation, the predicted strength of ER effects is of an order lower 
than the experimental results . On the other hand, the accurate 
theoretical models are usually too complex to use in dynamic simulation 
of ER fluids. Hence, 
a model which is easy to use but beyond the point-dipole results is 
needed in computer simulations.  

In this paper, we use a dipole-induced-dipole (DID) model \cite{yu} 
proposed by one of the authors to study the 
ground state of ER fluids in a wide range of the permittivity ratios: 
$0<\ep_p/\ep_f<\infty$. This model accounts for the multipolar 
interaction but is significantly simpler compared with existing 
multipolar theories \cite{cle,fri,fu}, therefore it can serve as a 
candidate for computer simulation instead of the traditional 
point-dipole approximation. Some computation has been carried out 
such as the calculation of the interacting force between particles of 
different sizes and various dielectric constants and the simulation of 
the athermal aggregation of particles in ER fluids, both in uniaxial 
and rotating fields \cite{yu,siu,wong}. The purpose of this paper is 
to  use this model to deal with structure formation in ER fluids and 
to offer some theoretical prediction as instructions to further dynamic 
simulations of ER fluids.  It is found the ground state of ER fluids 
may vary with the dielectric constant, and the critical dielectric 
contrast is estimated. Furthermore, we obtain the interaction between 
two chains containing particles of various dielectric constants and 
apply it to polydisperse ER fluids where the particles have the same 
size but different permittivities. The effects of polydispersity on 
structure formation are investigated. 

\section{Interaction between two particles}

We start with briefly reviewing the DID model and then applying it to 
deal with the interaction energy between two particles.
The DID model is defined from a multiple image method \cite{yu} which 
is based on a generalization of the image method to dielectric spheres. 
First consider a simple situation in which a point dipole $p$ is placed 
at a distance $r$ from the center of a sphere. The orientation of the 
dipole is perpendicular to the line joining the dipole and the center 
of the sphere. If the sphere is conducting, the induced image dipole 
is exactly given by $p'=-p(a/r)^3$ and at a distance $r'=a^2/r$ from 
the center. Generalizing this result to a sphere of dielectric constant 
$\ep_{p}$ placed in a host medium $\ep_f$ reads $p'=-\beta p (a/r)^3$ 
with $\beta$ as the dipolar factor 
$\beta=\frac{\ep_{p}-\ep_f}{\ep_{p}+2\ep_f}$. 
If the orientation of the point dipole is parallel to the axis, 
then $p'=2\beta p(a/r)^3$. In the limit $\beta \rightarrow 1$, 
the above results reduce to the conducting sphere case.

Then consider a pair of dielectric spheres A and B, of the same radius 
$a$ but different dielectric constants $\ep_{p1}$ and $\ep_{p2}$, 
separated by a distance $r$, in a base fluid 
of a dielectric constant $\ep_f$.  Upon the application of an electric 
field ${\bf E}_0$, the induced dipole moments in the individual spheres 
are respectively given by
\be
p_{a0}=\ep_f \beta E_0 a^3, p_{b0}=\ep_f \beta' E_0 a^3
\ee
with the dipole factors $\beta=\frac{\ep_{p1}-\ep_f}{\ep_{p1}+2\ep_f}$ 
and $\beta'=\frac{\ep_{p2}-\ep_f}{\ep_{p2}+2\ep_f}$. The initial dipole 
moment $p_{a0}$ induces an image dipole $p_{b1}$ in sphere B, 
while $p_{b1}$ induces another image dipole $p_{a2}$ in sphere A. 
As a result an infinite series of dipoles {$p_{a0}, p_{a1}, 
p_{a2},...$} are formed inside sphere A and the total dipole moments 
for transverse and longitudinal fields are respectively given by
\bee
p_{aL} &=& \sinh^3 \arf 
\sum_{n=1}^{\infty}[\frac{p_{a0}(2\beta)^{n-1}(2\beta')^{n-1}}{(\sinh 
n\arf+\sinh (n-1)\arf)^3}+
         \frac{p_{b0}a^3(2\beta)^{n}(2\beta')^{n-1}}{(r\sinh n\arf)^3}], 
\\ 
p_{aT} &=& \sinh^3 \arf 
\sum_{n=1}^{\infty}[\frac{p_{a0}(-\beta)^{n-1}(-\beta')^{n-1}}{(\sinh 
n\arf+\sinh (n-1)\arf)^3}+
         \frac{p_{b0}a^3(-\beta)^{n}(-\beta')^{n-1}}{(r\sinh n\arf)^3}].
\eee
The parameter $\arf$ in Eq. (1) and (2) satisfies
\be
\cosh\arf=\frac{r^2-2a^2}{2a^2}.
\ee
Similar expressions for the total dipole moments inside sphere B, namely 
$p_{bL}$ and $p_{bT}$, can be obtained by 
interchanging $\beta$ and $\beta'$.

We should remark that the present generalization is only approximate 
because there is no image method for a dielectric sphere. An integral 
equation approach was proposed to examine the validity of the multiple 
image method in Ref. \cite{yu}. It was shown that Eq. (2) and (3) can 
produce good results at high dielectric contrast $\beta\rightarrow1$ 
as expected, and the model keeps reasonable even in the low contrast 
case \cite{yu,siu}.

How many terms are retained in the analytic multiple image result 
determines different models. With only the $n=1$ term in the series of 
Eq. (2) and (3), it reduces to the traditional point-dipole 
approximation, while the DID model is defined by retaining the first 
two terms ($n=1$ and $n=2$). Taking part of the multipole effects into 
account, the DID model is generally better than the PD results, 
especially in polydisperse systems, and can be chosen as an effective 
approximation to calculate multipolar interaction \cite{yu,siu,wong}. 

The incorporation of multipole effects leads to one of the obvious 
differences between the PD and DID model: the symmetry of negative and 
positive $\beta$s will be broken in the latter, which is illustrated 
in Fig. 1. This asymmetry of the dipolar factor will also be exhibited 
in the interaction energy between particles, as shown below, and finally 
affect the structure formation in ER fluids. It implies that different 
structures may form when the dielectric constant of particles is smaller 
than that of the host fluid.

Now we begin to apply the DID model to deal with the interaction energy 
between two particles. The electrostatic energy of the two particles 
upon the application of an electric field ${\bf E}_0$ at an arbitrary 
angle $\theta$ (Fig. 2a) is given by \cite{jac}
\be
W=-\frac{E_0}{2}[(p_{aL}+p_{bL})\cos^2\theta+(p_{aT}+p_{bT})\sin^2\theta].
\ee
Substituting the dipole moment expressions into Eq. (5) and letting 
$\lambda=\beta'/\beta$, $p_0=p_{a0}$, we can get the energy using the 
DID approximation as
\bee
W&=&-\frac{1+\lambda}{2}E_0p_0+\lambda(-\frac{p_0^2}{\ep_f}\frac{3\cos^2\theta 
-1}{r^3})
  +\frac{\lambda+\lambda^2}{2}[-\frac{p_0^2}{\ep_f 
a^3}\beta\mu(3\cos^2\theta+1)]\nonumber\\
   &+&\lambda^2[-\frac{p_0^2}{\ep_f 
r^3}\beta^2\nu(9\cos^2\theta-1)]\nonumber\\
   &\equiv&W_f+W_0+W_1+W_2,
\eee
where the dimensionless parameters have been defined:
\be
\mu=(\frac{1}{r'^2-1})^3,\ \
\nu=(\frac{1}{r'^2-2})^3,\ \
r'\equiv\frac{r}{a}.\nonumber
\ee
The first term $W_f$ in Eq. (6) is the energy of the individual particles 
by the applied field, which is independent of the relative position of 
the particles, while the remaining terms correspond to the interaction 
energy which determines the formation of ground states \cite{Tao1}.

Note that the term $W_0$ is, if $\lambda=1$, exactly the result of 
Tao {\em et al.}'s work \cite{Tao1} , in which they dealt with the 
monodisperse case in the use of a PD approximation and concluded that 
a bct lattice will form as the ground state. 
So what we are interested in is the last two terms which describe the 
DID correction to the total interparticle energy.
Adding the hard-core repulsion which keeps the particles from coalescing 
($r'>2$), $W_1$ and $W_2$ can be rewritten as
\bee
W_1 &=&- \beta \frac{\lambda+\lambda^2}{2} \sum_{n=6}' a_{\frac{n-6}{2}} 
\frac{3\cos^2 \theta +1}{r'^n}
=- \beta \frac{\lambda+\lambda^2}{2} \sum_{n=6}' a_{\frac{n-6}{2}}[4+ 
\frac{3}{n}\rho \frac{\partial}{\partial \rho}] \frac{1}{r'^n}\\
W_2&=&-\beta^2 \lambda^2 \sum_{n=9}'a_{\frac{n-9}{2}} \frac{9\cos^2 
\theta -1}{r'^n}
=-\beta^2 \lambda^2 \sum_{n=9}'a_{\frac{n-9}{2}}[8+ \frac{9}{n}\rho 
\frac{\partial}{\partial \rho}] \frac{1}{r'^n}, 
\eee
Here we have taken $\frac{p_0^2}{\ep_f a^3}$ as the energy unit. 
$\sum_{n=i}'(...)$ means the summation over $n=i, i+2, i+4,...$, 
the coefficient $a_i=\frac{3 (3+1) ... (3+i-1)}{i!}$, and 
$\rho=[(x_z-x_b)^2+(y_a-y_b)^2]^{1/2}=\sqrt{r^2-(z_a-z_b)^2}$ 
($(x_i, y_i, z_i)$ with $i=a,b$ being the coordinates of the particles 
and $z$ axis is chosen as along the field direction). 
Obviously, the DID correction cannot be ignored when $\beta$ is large, 
while the term $W_1$ being proportional to $\beta$ instead of 
$\vert \beta \vert$ shows the asymmetry of $\beta$ in the DID model.

\section{Monodisperse systems}

Now we study a monodisperse ER fluid composed of particles with a 
dielectric constant $\epsilon_p$, which can be larger ($\beta>0$) 
or smaller ($\beta<0$) than that of the fluid $\epsilon_f$. 
It has been mentioned that the interacting particles in a ER fluid 
will first form chains between electrode plates, then the chains 
aggregate into columns containing microstructures. 
Combined with the images, the chains can be treated as infinite 
if the distance between two plates is large enough \cite{Tao1}. 
The interaction energy per particle is divided into two parts: 
one is from the self-energy of an infinite chain, i.e., the interaction 
energy between the particles belonging to the same chain; 
the other from interaction between different chains \cite{Tao1}. 

\subsection{$u^s$, the self- energy}

Consider an infinite chain containing particles of radius $a$ and 
dielectric constant $\ep_{p}$
at ${\bf r_j}=2aj\hat{{\bf z}}$ ($j=0, \pm 1, \pm 2,...$) (Fig. 2b). 
The self-energy of the chain is given by
\be
u^s=\frac{1}{2} \sum_{j\neq 0}W_i({\bf r_j})=\frac{1}{2} \sum_{j\neq 
0}W_i(r=2aj,\theta=0).
\ee
$W_i$ refers to the interaction energy between two particles.  
Substituting the DID terms in the particle interaction, 
$W_1$ and $W_2$, into Eq. (10), we get
\bee
u_1^s &=& -\beta \sum_{n=6}' \frac{a_{\frac{n-6}{2}}}{2^{n-2}} 
\zeta(n)=-0.149449\beta, \\
u_2^s &=& -\beta^2 \sum_{n=9}' 
\frac{a_{\frac{n-9}{2}}}{2^{\frac{n+3}{2}}} \zeta(n)=-0.125047\beta^2,
\eee
where $\zeta(n)$ is the Zeta function defined as 
$\zeta(n)=\sum_{n=1}^{\infty} \frac{1}{j^n}$. Again, the DID correction 
is comparable to the PD result \cite{Tao1}, $-0.300514$, when $\beta$ is 
large. 
The total self-energy $u^s$ in a DID model is the sum of $u_1^s$, 
$u_2^s$  and that from PD assumption
\be
u^s=-0.300514-0.149449\beta-0.125047\beta^2,
\ee
which represents a correction to the established PD results.

\subsection{$u^i$, the interaction energy}

The interchain energy between two parallel infinite chains, 
with vertical shift $z$ and separated by distance $\rho$,
is given by $\frac{1}{2} u^i(\rho, z)$, where $u^i(\rho, z)$ is the 
interaction between one dielectric particle at 
${\bf r}_p={\bf \rho}+z \hat{{\bf z}}$ near the infinite chain 
in which particles locate at ${\bf r_j}=
2aj\hat{{\bf z}}$ ($j=0, \pm 1, \pm 2,...$) (Fig. 2c),

\be
u^i(\rho, z) = \sum_{j} W_i ({\bf r}_j-{\bf r}_p).
\ee 
Hence we have the DID correction of $u^i$

\be
u_1^i(\rho, z)=- \beta \sum_{n=6}' a_{\frac{n-6}{2}}[4+ \frac{3}{n}\rho 
\frac{\partial}{\partial \rho}] 
\sum_{j=-\infty}^{+\infty}\frac{1}{[\rho^2+(z-2aj)^2]^{n/2}}
\ee
and
\be
u_2^i(\rho, z)= -\beta^2 \sum_{n=9}'a_{\frac{n-9}{2}}[8+ \frac{9}{n}\rho 
\frac{\partial}{\partial \rho}] 
 \sum_{j=-\infty}^{+\infty}\frac{1}{[\rho^2+(z-2aj)^2]^{n/2}}.
\ee
Following a Fourier series technique proposed by Tao {\em et al.} 
\cite{Tao1} we expand $u_1^i$ and $u_2^i$ into
\bee
u_1^i(\rho,z)&=& - \sqrt{\pi} \beta  \sum_{n=6}' a_{\frac{n-6}{2}}
  (\frac{a}{\rho})^{n-1}\{ \frac{n+3}{2n} \frac{\Gamma 
(\frac{n-1}{2})}{\Gamma (\frac{n}{2})} + 
  \frac{S_1}{n\Gamma (\frac{n}{2})} \}\\
u_2^i(\rho,z)&=&- \sqrt{\pi} \beta^2  \sum_{n=9}'a_{\frac{n-9}{2}} 
2^{\frac{n-9}{2}}
 (\frac{a}{\rho})^{n-1}\{ \frac{-n+9}{2n} \frac{\Gamma 
(\frac{n-1}{2})}{\Gamma (\frac{n}{2})} + 
  \frac{S_2}{n\Gamma (\frac{n}{2})} \}
\eee
with
\bee
S_1 &=& \sum_{s=1}^{\infty} (\frac{\rho \omega}{2})^{\frac{n-1}{2}} 
[(3+5n)K_{\frac{n-1}{2}}(\rho \omega)
  -3 \rho \omega K_{\frac{n+1}{2}}(\rho \omega) - 3 \rho \omega 
K_{\frac{n-3}{2}}(\rho \omega)]\cos (\frac{s\pi z}{a}),\\
S_2 &=& \sum_{s=1}^{\infty} (\frac{\rho \omega}{2})^{\frac{n-1}{2}} 
[(9+7n)K_{\frac{n-1}{2}}(\rho \omega)
  -9 \rho \omega K_{\frac{n+1}{2}}(\rho \omega) - 9 \rho \omega 
K_{\frac{n-3}{2}}(\rho \omega)]\cos (\frac{s\pi z}{a}),\\
\omega &\equiv& s\pi/a.
\eee 
$\Gamma(x)$ and $K_i(x)$ in the above equations are the gamma function 
and i-th order modified Bessel function, respectively.
The sums above can be easily evaluated numerically, and the expression 
of $u^i(\rho,z)$ is given by 
$u^i(\rho, z)=u^i_0(\rho, z)+u^i_1(\rho, z)+ u^i_2(\rho, z)$, 
where $u^i_0(\rho, z)$ is the PD result as \cite{Tao1}
 \be
u_0^i(\rho,z)=\lambda \sum_{s=1}^{\infty} 2 \pi^2 s^2 K_0(\frac{s\pi 
\rho}{a})\cos (\frac{s\pi z}{a}).
 \ee

\subsection{Possible ground states}

The total interaction energy per particle in a certain configuration is 
\cite{tao98}
\be
u=u^s+\frac{1}{2}\sum_{k}'u^i(\rho_k,z_k),
\ee
where $\sum_k'$ denotes the summation over all chains labeled $k$ except 
the one containing the considered particle. Since
the self-energy of a chain is independent of the structure, what really 
affects the energy difference between various 
lattices is $u^i(\rho, z)$. Fig. 3 shows the dependence of $u^i$ on the 
shift $z$ for different values of $\rho$ when 
$\beta>0$. The results from the PD approximation are also plotted for 
comparison. The interaction may be either 
attractive or repulsive depending on the shift $z$, and the range of $z$ 
in which  
two chains attract each other  enlarges quickly when $\beta$ increases. 
This implies
a tendency to form a more close packed structure than the bct lattice 
with increasing dielectric contract.
 The most possible candidate is the fcc lattice, which is 
nearest to the bct one in energy \cite{Tao1}. An estimation including the 
nearest and the next-nearest neighboring chains can give 
the energy gap between the two lattices

\bee
\Delta u & \equiv & u_{fcc}-u_{bct} \doteq u^i(\rho=2a, 
z=0)-2u^i(\rho=\sqrt{6}a, z=0)\nonumber\\
&=&0.0110-0.0230\beta+0.0127 \beta^2,
\eee
which decreases when $\beta$ increases from 0 to 1. For the limit case 
$\beta=1$ ($\ep_p\gg \ep_f$), $\Delta u \simeq 6.6
\times 10^{-4}$ and the two phases can be regarded as degenerate, in 
agreement with the conclusion of previous work \cite{dav,cle}.

Reversed situation happens when $\beta<0$ (Fig. 4 ). 
Now it's the repulsive region that enlarges and the system tends to form 
a looser structure as $\beta$ approaches the negative limit 
$-\frac{1}{2}$ ($\ep_p \ll \ep_f$). 
The loosest structure is certainly the one composed of separate chains. 
Calculating the difference between the bct lattice and separate chains, 
we get

\bee
\Delta u & \doteq & 2u^i(\rho=\sqrt{3}a, z=a)+2u^i(\rho=\sqrt{6}a, 
z=0)\nonumber\\
&=&-0.081-0.300\beta-0.155 \beta^2.
\eee 
When $\beta=-0.323$ a transition from the bct lattice to separate chains 
happens and the latter becomes more and more stable 
while the absolute value of $\beta$ increases.

Finally we want to make a comparison with Clercx {\em et al.}'s elegant 
work based on a multipole-expansion theory \cite{cle}. 
They expanded the potentials inside and outside the spheres in terms of 
solid spherical harmonics and gained the values of multipole moments 
$Q^{i}_{lm}$ by solving a set of linear equations determined by the 
boundary conditions at the grain surfaces. The upper limit of $l$, 
denoted by $L$, determines how many multipole effects are considered: 
$L=1$ refers to the simple dipole approximation and $L=\infty$ to 
the exact calculation where all multipole effects are included. 
This multipole expansion theory and our DID model are based on different 
pictures and there exists no direct and exact equivalent relationship 
between them. The calculation of the interparticle energy $W_i$ using 
Eq. (6) and Clercx {\em et al.}'s model shows that the DID results 
reflects, although not exactly, some characteristics of the third 
order($L=3$) multipole-expansion theory, i.e., an octupole effect. 
This is particularly obvious when $\theta=\pi/2$ even in the 
touching-particle case (See Fig. 5), where the point-dipole approximation 
is known to err considerably.  We expect that the DID results may be 
better than the first approximation of Clercx {\em et al.}'s model 
since the former takes higher multipole effects into consideration. 
Unfortunately the available data related with the energy ground state 
are limited in Ref. \cite{cle} and we cannot give a thorough comparison. 
However, from Table 1 in which the ratios of the total induced dipole 
per particle between the bcc and the fcc lattice calculated with our 
model are compared with the available results of Clercx {\em et al.}, 
we can still see that  the DID results is closer to the exact solutions 
than the dipole approximation in a wide range of dielectric mismatch.

\section{polydisperse systems}

So far we have discussed the interaction energy and ground states in 
an ER fluid containing the same particles. Now we begin to investigate 
a more "natural" polydisperse system in which the particle dielectric 
constant $\epsilon_p$, and consequently the dipolar factor $\beta$, 
has a probability distribution. Let the average interaction energy 
between a pair of particles in this system denoted by $W^*_i$. 
Assuming the dipolar factors $\beta$ and $\beta'$ of the two particles 
are independent random variables which have the same distribution

\be
P(\beta)=\frac{1}{\sqrt{2\pi}\sigma}\exp(-\frac{(\beta-\beta_0)^2}{2\sigma^2}),
\ee
we can get $W^*_i$ by taking an average of Eq.(6) over $\beta$ and 
$\beta'$ as

\be
W^*_i=\overline{W_i}+W_{\delta},
\ee
where $\overline{W_i}$ is the interaction energy, including the PD 
results and two DID terms, between identical particles
with the average dipolar factor $\beta_0$, while $W_{\delta}$ is the 
polydispersity correction

\be
W_{\delta}=-\epsilon_f E_0^2 a^3 \{ \sigma^2 \beta_0 \mu (3 \cos^2 \theta 
+1)+\nu (2 \beta_0^2 \sigma^2 +\sigma^4)(9 \cos^2 
\theta -1)\}
\ee 

with $\sigma$ and $\beta_0$ as the standard deviation and mean of the 
distribution of $\beta$.
On the basis of the expression of interparticle energy, we can easily 
obtain the average self-energy of an infinite
chain $u^{s*}$ and the interaction energy between two chains $u^{i*}$, 
in a polydisperse system where particle permittivities are distributed 
randomly, as

\bee
u^{s*}&=& \overline{u^s} +u^s_{\delta},\\
u^{i*}(\rho, z)&=&\overline{u^i}(\rho, z) + u^i_{\delta}(\rho, z),
\eee
where the energy increment caused by polydispersity $u^s_{\delta}$ 
and $u^i_{\delta}$ are given by

\bee
u^s_{\delta}&=&\delta^2 \overline{u^s_1}+ (2 \delta^2 +\delta^4) 
\overline{u^s_2}\\
u^i_{\delta}&=&\delta^2 \overline{u^i_1}+ (2 \delta^2 +\delta^4) 
\overline{u^i_2}
\eee
In the above equations, $\delta$ is defined as $\sigma/\beta_0$ to 
describe the degree of polydispersity, $\overline{u^s}$,
$\overline{u^i}$ are the total self-energy and interchain energy for 
a monodisperse system in which all particles have the same dipolar 
factor $\beta_0$, while $\overline{u^s_1}$,
$\overline{u^s_2}$ and $\overline{u^i_1}$,
$\overline{u^i_2}$ are the corresponding DID corrections. 
They can be calculated using Eq. (10) - (22).
 
As a consequence of the change of the chain energies caused by 
polydispersity, the energy gaps between different lattices
also vary and hence affect structure formation in ER fluids. 
Fig. 6 shows both $\Delta \overline{u} -\beta_0$ and 
$\Delta u_{\delta} - \beta_0$ with $\delta=0.1$ to exhibit the effect 
of polydispersity on ground states. $\Delta \overline{u}$ 
is defined as $ \overline{u_{bct}}-\overline{u_{fcc}}$
when $\beta_0>0$ and $\overline{u_{bct}}-\overline{u_{sc}}$ 
when $\beta_0<0$ (Here "sc" means separated chains) for a system 
composed of particles $\beta_0$, while $\Delta u_{\delta}$ is that 
caused by polydispersity. It is shown that $\Delta u_{\delta}$ 
keeps opposite to $\Delta \overline{u}$ when $\beta_0$ is larger than 
$-0.323$ and not very near $1.0$. In this range the bct
lattice has been proved to be the stable ground state in a monodisperse 
system, and the existence of polydispersity tends to 
destroy its stableness. However, the magnitude of the ratio $\Delta 
u_{\delta} / \Delta \overline{u}$ is only of the order
$10^{-2}$, so we can conclude that, in a polydisperse system where the 
dielectric mismatch between particles and the fluid
is not very large,  similar structures will form but the ground state 
is not so stable as that in a monodisperse system.

All the results discussed above are based on the assumption that the 
dielectric constants of particles are distributed
at random in the configuration. But the question is: Is the distribution 
really random, especially when the dielectric 
constants of particles differ greatly from each other? In the extreme 
case, will the phenomenon "phase separation" happen,
i.e., identical particles gather together upon the application of the 
electric field? In order to  investigate this
problem, we consider a two-component fluid containing particles with 
$\beta_1$ and $\beta_2$. The fractions of particles are
 supposed to be $p_1$ and $p_2$, respectively. First assume a totally 
random configuration will form in this system. The average 
 self-energy $u^s_{mix}$ and interchain energy $u^i_{mix}$ in such a 
system can be calculated by means of the method described according to
the binary distribution

\be
P(\beta)=p_1 \delta(\beta-\beta_1)+p_2 \delta(\beta-\beta_2).
\ee
Then consider another configuration in which only identical particles 
aggregate into "uniform" chains. Two types of chains 
are formed in this system: one, called chain A, is composed of particles 
$\beta_1$, and another, chain B, of $\beta_2$. The
average self-energy $u^s_{unif}$ is determined by

\be
u^s_{unif}=p_1 u^s\mid_{\beta=\beta1}+p_2 u^s\mid_{\beta=\beta_2},
\ee
and the interaction energy between two uniform chains satisfies
\be
u^i_{unif}(\rho, z)=u^i_0(\rho, z)+u^i_1(\rho, z)+u^i_2(\rho,z)
\ee
with
\bee
u_0^i(\rho,z)&=&\lambda \sum_{s=1}^{\infty} 2 \pi^2 s^2 K_0(\frac{s\pi 
\rho}{a})\cos (\frac{s\pi z}{a}),\\
u_1^i(\rho,z)&=& - \sqrt{\pi} \beta \frac{\lambda+\lambda^2}{2} 
\sum_{n=6}' a_{\frac{n-6}{2}}
  (\frac{1}{\rho'})^{n-1}\{ \frac{n+3}{2n} \frac{\Gamma 
(\frac{n-1}{2})}{\Gamma (\frac{n}{2})} + 
  \frac{S_1}{n\Gamma (\frac{n}{2})} \},\\
u_2^i(\rho,z)&=&- \sqrt{\pi} \beta^2 \lambda^2 
\sum_{n=9}'a_{\frac{n-9}{2}} 2^{\frac{n-9}{2}}
 (\frac{1}{\rho'})^{n-1}\{ \frac{-n+9}{2n} \frac{\Gamma 
(\frac{n-1}{2})}{\Gamma (\frac{n}{2})} + 
  \frac{S_2}{n\Gamma (\frac{n}{2})} \}.
\eee
$\lambda$ is the ratio of $\beta$s of the two chains and other parameters 
are defined as previously.

We have compared the energies in these two configurations and plotted the 
results in Fig. 7. The upper graph
in Fig. 7 shows the dependence of $\Delta u^s \equiv u^s_{mix}- 
u^s_{unif}$ on $\beta_1$ and $\beta_2$ , and the lower one
shows $\Delta u^i \equiv u^i_{mix}- u^i_{unif}$. Note that here 
$u^i_{unif}$ is the interaction between chain A and chain B.
The results when we select other chain configuration such as chains AA or 
BB are similar. The fraction of the particles 
with $\beta_1$ is 0.5, and the relative coordinates of the interacting 
chains are chosen as  $\rho = \sqrt{3}a$, $z=a$.
It is shown that $u^s_{mis}$ is always larger than
 $u^s_{unif}$, 
while $\Delta u^i$ is usually negative for most  values of $\beta_1$ and 
$\beta_2$ but much smaller than $\Delta u^s$ in
magnitude.  Since the self-energy is usually dominant in the total 
interaction energy per particle, chains of identical 
particles may be more stable than those containing different types of 
particles, 
particularly when $\beta_1$ and $\beta_2$ approach the positive limit 
1.0 and negative limit -0.5 simultaneously.

The interaction form in polydisperse fluids is more complex than what 
we have considered because it is sensitive to the micro structure 
formed in the ER solid and the number of possible configurations in 
a polydisperse system is much larger than that in a monodisperse one. 
The energy gap between the random and phase separation configurations 
may be narrowed because of the contribution of interaction between 
chains. But we can still expect that the ground state of such a 
two-component system may contain quite a few of these "uniform" chains 
instead of a totally random configuration. And the simplicity of our 
DID model will also make it possible to carry out computer simulations 
in polydisperse systems and study the dielectric effects in a more 
detailed way \cite{wong}.

\section{Conclusions}

In conclusion, we have presented a DID model to study the dielectric 
effects on structure formation in ER fluids. Based on the DID 
expression of the interacting energy of two particles, 
we have corrected the PD results of the self-energy and the interaction 
energy of chains, including multipole effects partially. 
Series expressions of these energies are obtained and used to calculate 
the effects of dielectric mismatch between particles and the fluid. 
Both monodisperse and polydisperse cases are discussed and some 
interesting results are obtained as follows.

(1) In monodisperse systems where particles are negatively polarized, 
there may exist a phase transition form the bct lattice
to the configuration of separate chains when $\beta<-0.323$;

(2) Polydispersity of particle dielectric constants in ER fluids will 
cause the bct ground state not so stable as that
in monodisperse systems when the dielectric mismatch between the particle 
and the fluid is not very large;

(3) When the dielectric constants of particles differ much with each 
other, identical particles tend to aggregate into 
uniform chains.

The ER effects in polydisperse system will be studied more deeply in 
our future simulation work based on the DID model.

\section{Acknowledgment}

This work was supported in part by the Direct Grant for Research and 
in part by the RGC Earmarked Grant. We acknowledge useful discussion 
with Professor Z. Y. Li.

\begin{figure}[h]
\caption{The total induced dipole inside a couple of identical spheres 
A and B upon the application of ${\bf E}_0$ in the DID model when the 
dipolar factor $\beta>0$ and $\beta<0$.}
\end{figure}

\begin{figure}[h]
\caption{(a) two interacting particles; (b) the self-energy $u^s$ of 
an infinite chain; (c) interaction energy $u^i$ between an infinite 
chain and a neighboring particle at $(\bf{\rho}, z)$.}
\end{figure}

\begin{figure}[h]
\caption{The dependence of $u^i(\rho,z)$ 
(in units of $\frac{p_0^2}{\ep_f a^3}$) on z for different positive  
$\beta$s in the monodisperse system. Solid lines: $\rho=2a$; 
Dashed lines: $\rho=\sqrt{5}a$; Dotted lines: $\rho=\sqrt{6}a$.} 
\end{figure}

\begin{figure}[h]
\caption{The dependence of $u^i(\rho,z)$ 
(in units of $\frac{p_0^2}{\ep_f a^3}$) on z for different negative 
$\beta$s in the monodisperse system. Solid lines:
$\rho=2a$; Dashed lines: $\rho=\sqrt{5}a$; Dotted lines: $\rho=\sqrt{6}a$}
\end{figure}

\begin{figure}[h]
\caption{The relative interaction energy
(in units of $\frac{p_0^2}{\ep_f a^3}$) between a couple of identical 
particles vs the dipolar factor $\beta$ when $r=2$ and $\theta=\pi/2$. 
Solid line: the DID model; Dashed line: the multipole expansion theory 
with $L=3$;  Dotted line: the multipole expansion theory with $L=1$.}
\end{figure}

\begin{figure}[h]
\caption{(a) The energy gap
(in units of $\frac{p_0^2}{\ep_f a^3}$) between the bct and fcc lattices 
$\Delta \overline{u}$ for a monodisperse system of particles 
$\beta_0>0$ (solid line) and the corresponding polydispersity 
correction   $\Delta u_{\delta}$  (dashed lines) ;
 (b) The energy gap between the bct lattice and separate chains 
$\Delta \overline{u}$ for a monodisperse system of particles 
$\beta_0<0$ (solid line) and the corresponding polydispersity 
correction   $\Delta u_{\delta}$  (dashed lines) .}
\end{figure}

\begin{figure}[h]
\caption{The difference of the self-energy $\Delta u^s$ (the upper one) 
and the interchain energy $\Delta u^i$ (the lower one) 
between a random configuration and a system containing only uniform 
chains. All energies are in units of $\frac{p_0^2}{\ep_f a^3}$.}
\end{figure}

\newpage

\centerline{\epsfig{file=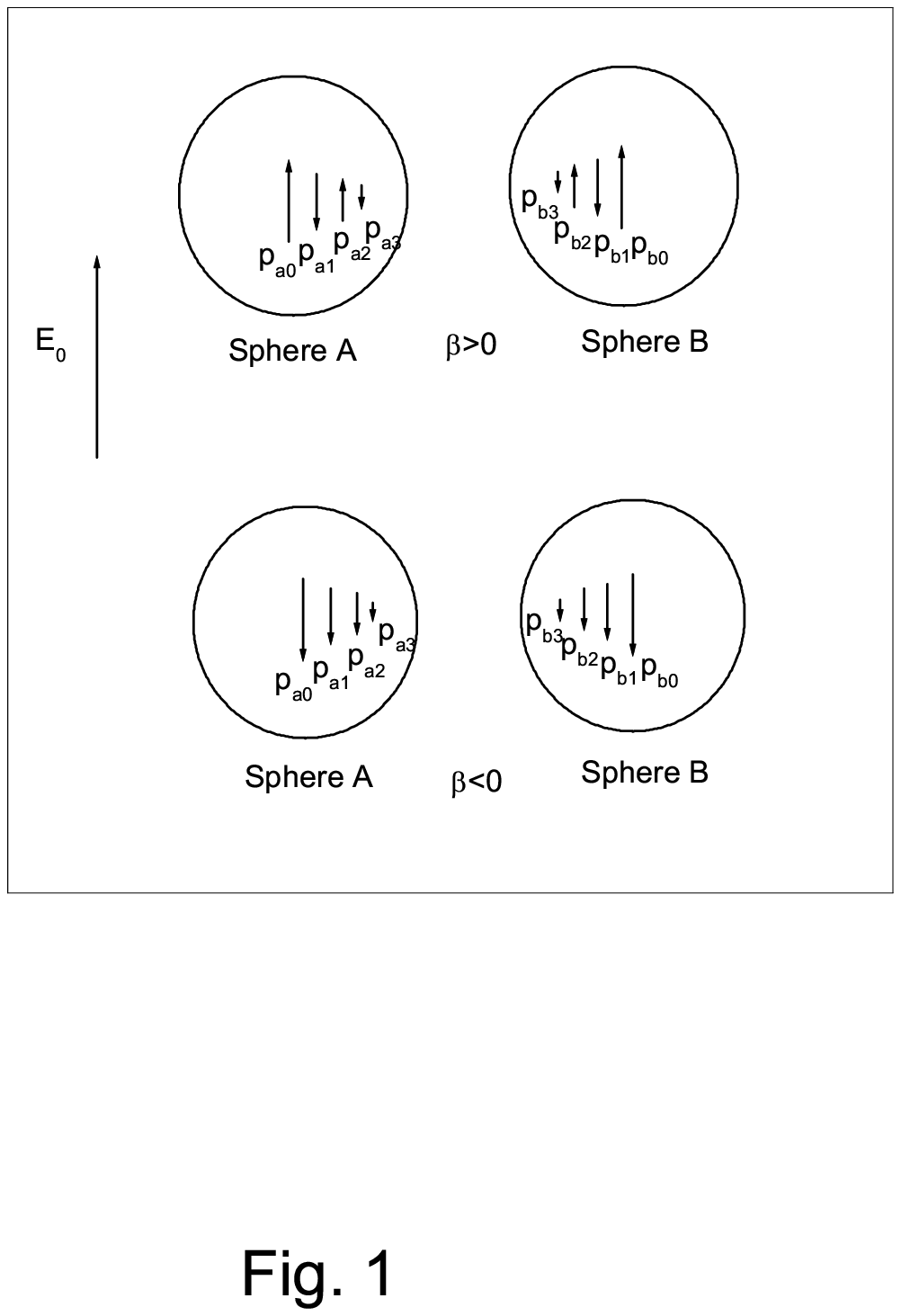,width=600pt}}

\centerline{\epsfig{file=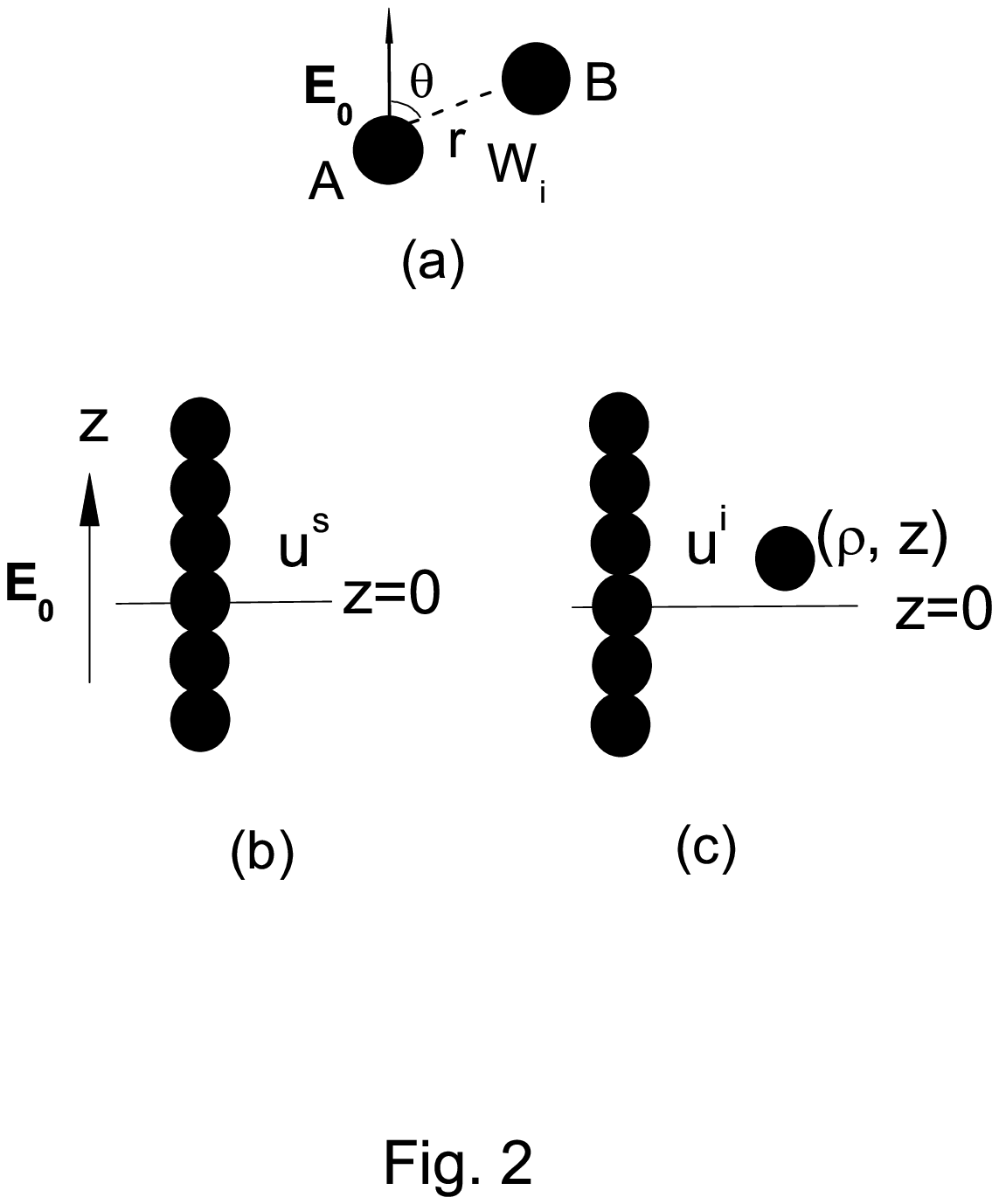,width=600pt}}

\centerline{\epsfig{file=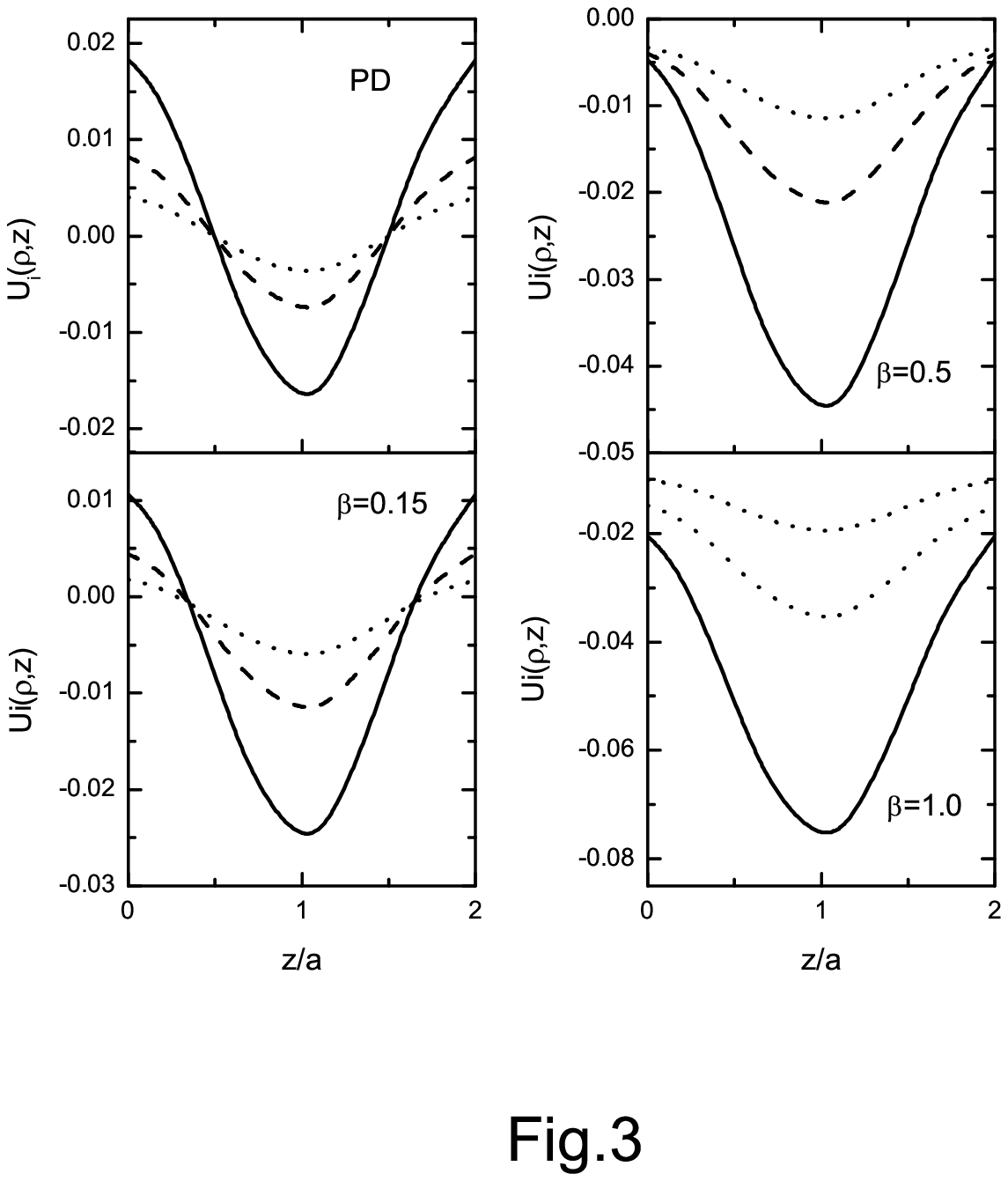,width=600pt}}

\centerline{\epsfig{file=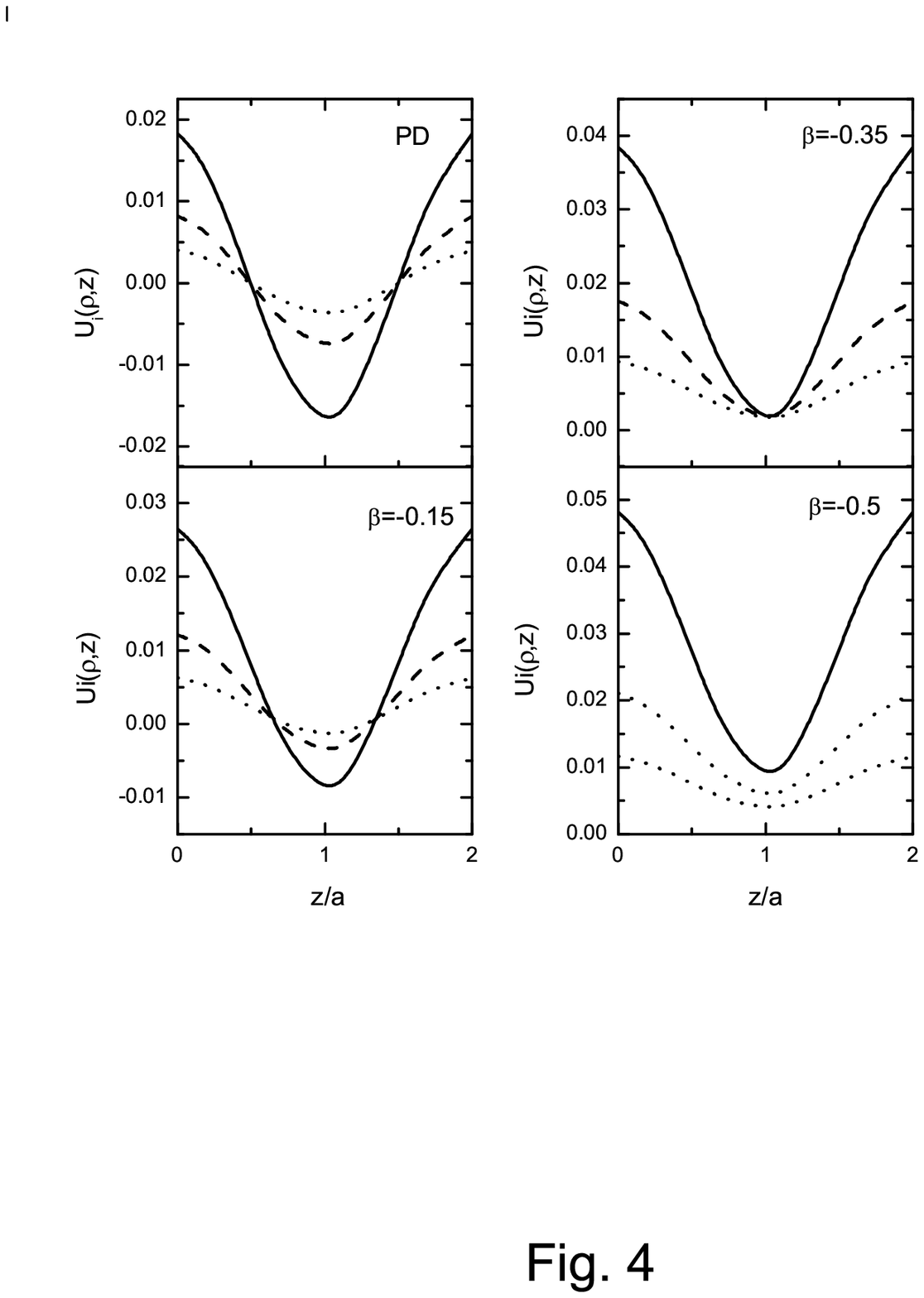,width=600pt}}

\centerline{\epsfig{file=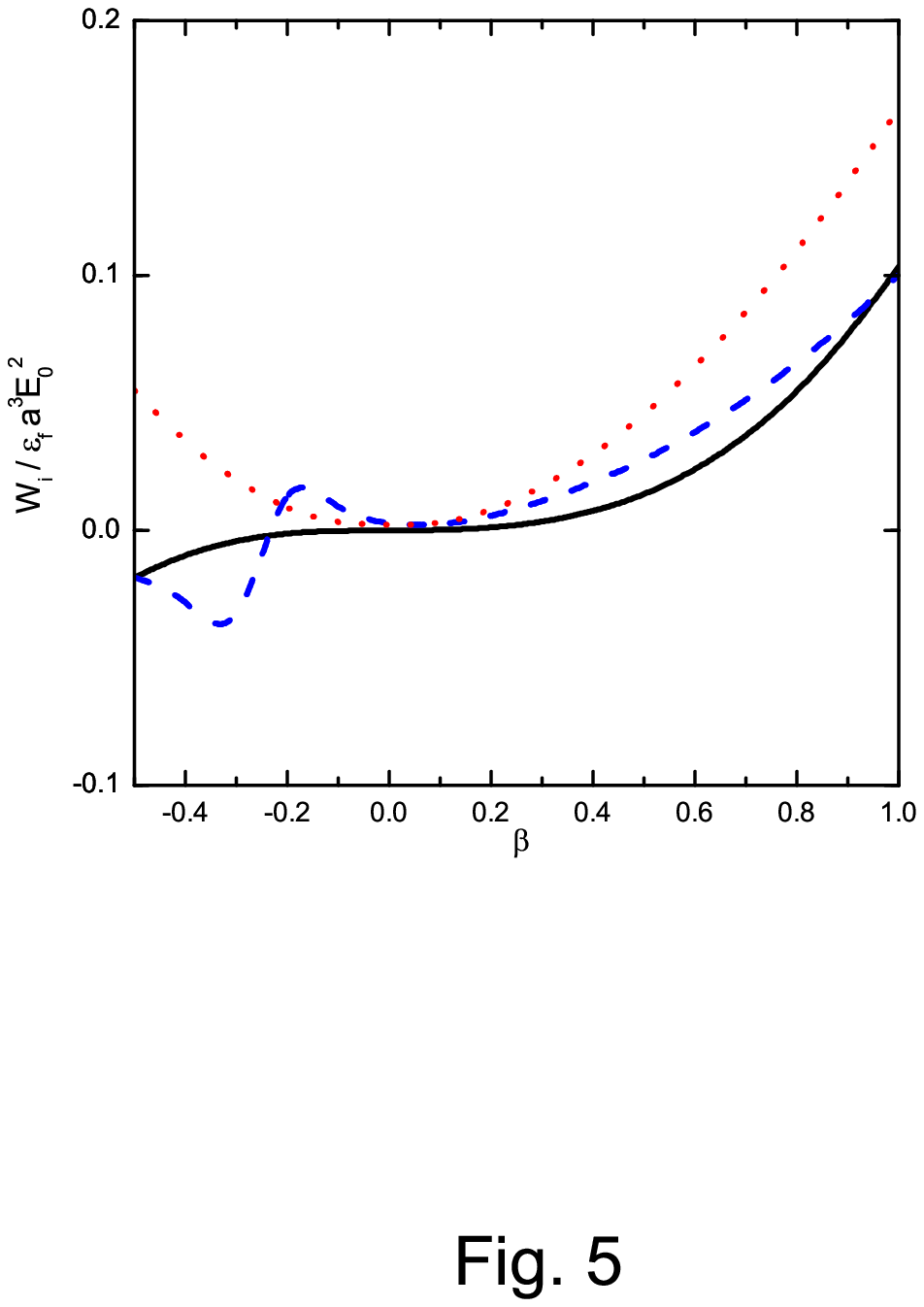,width=600pt}}

\centerline{\epsfig{file=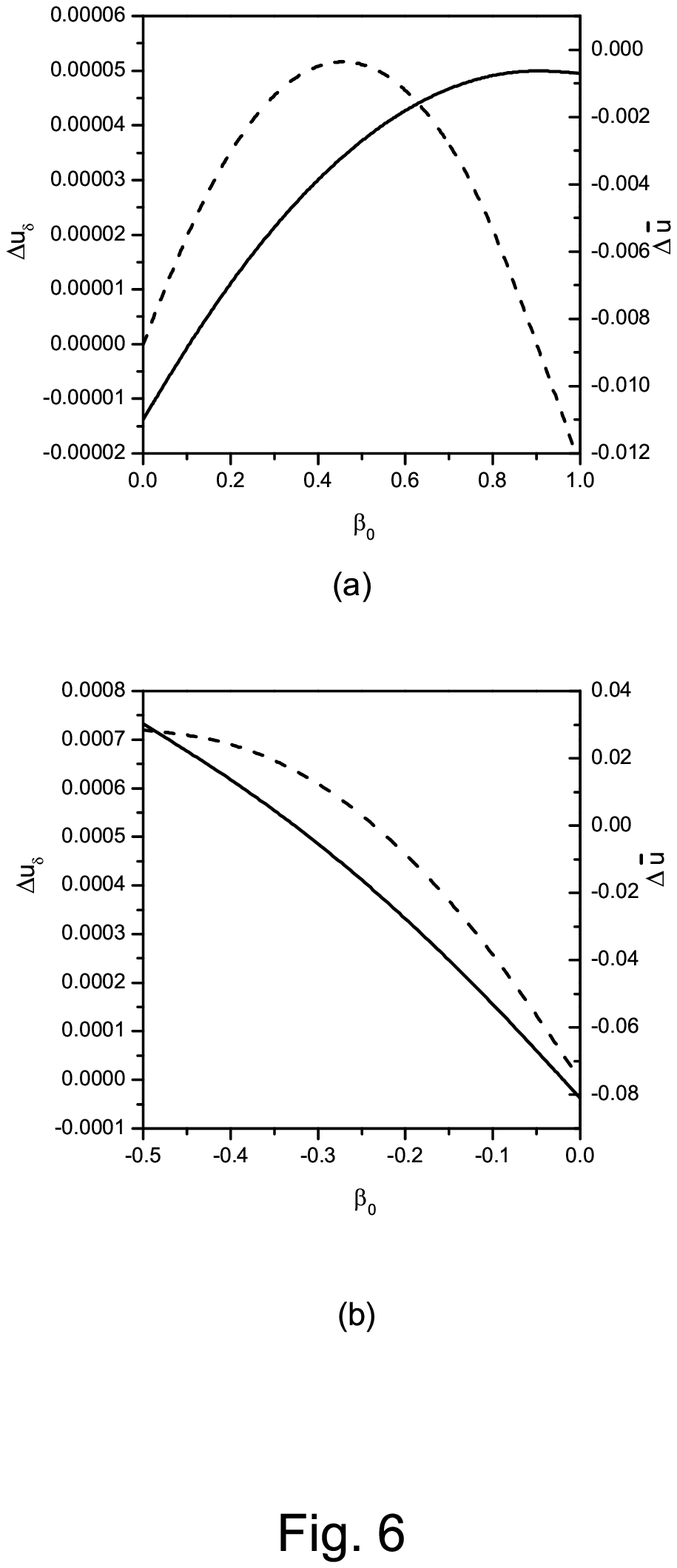,width=500pt}}

\centerline{\epsfig{file=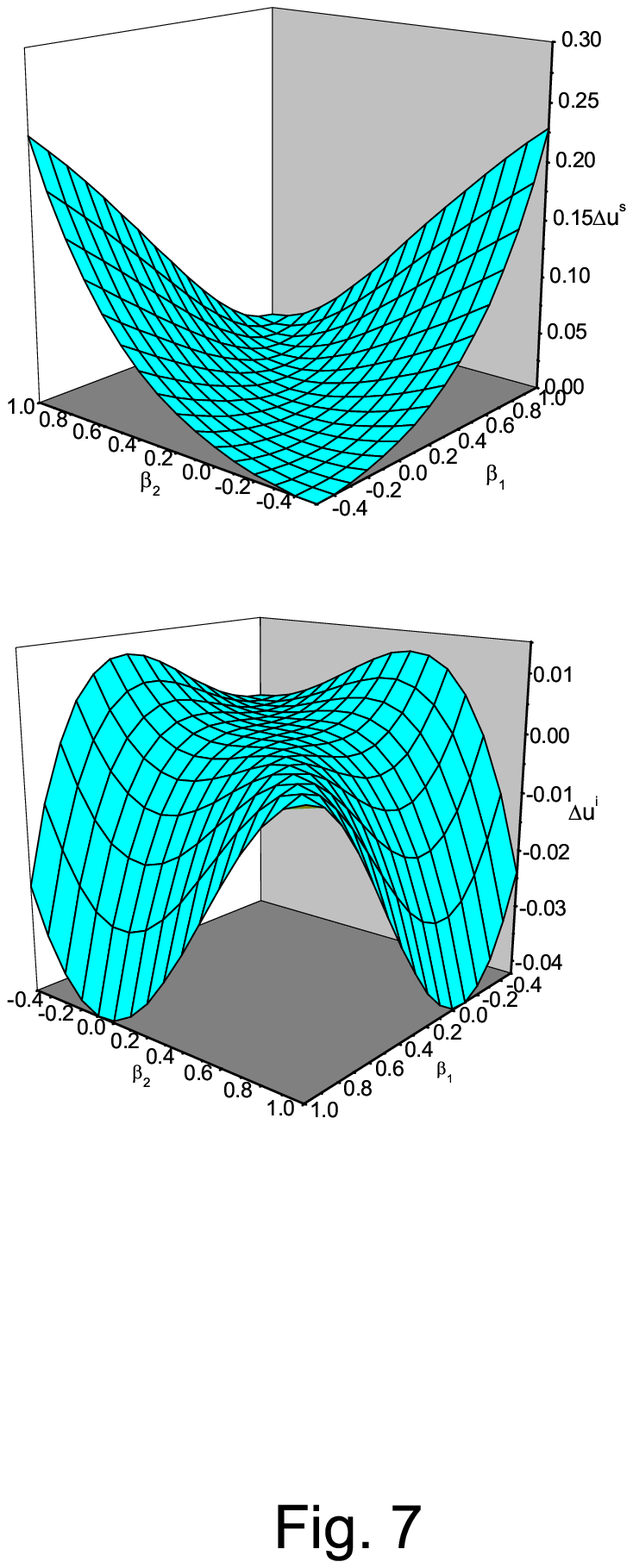,width=500pt}}

\begin{table}

\caption{The comparison between the DID model and the full multipole 
theory proposed by Clercx and Bossis . 
$\alpha$
is defined as $\epsilon_p/\epsilon_f$, where $\epsilon_p$ and 
$\epsilon_f$ are the dielectric constants of the particles 
and
the fluid, respectively. $\Delta_{DID}$ is the polarization ratio between 
the bct and fcc lattices using the present model. 
$\Delta_1$ and $\Delta_{\infty}$
are from the first-order approximation and the exact results of the full 
multipole theory, respectively.}

\begin{center}
\begin{tabular}{rrrr} \hline
 & $\Delta_{DID}$ & $ \Delta_1$ & $\Delta_{\infty}$\\\hline
 $\arf=0$ & 0.984 & 0.992 & 0.979\\ \hline
 $\arf=10$ & 1.001 & 1.039 &1.010\\\hline
 $\arf=100$ & 1.000 & 1.082 & 1.008\\\hline
\end{tabular}
\end{center}
\end{table}

\end{document}